\documentclass[preprint]{aastex}

\newcommand{\kms}{km~s$^{-1}$}
\newcommand{\myr}{M$_\odot$~yr$^{-1}$}
\newcommand{\whz}{W~Hz$^{-1}$}

\slugcomment{To appear in the {\it Astronomical Journal}}

\begin{document}

\title{A Deep VLA Radio Continuum Survey of the Core and Outskirts of the Coma Cluster}

\author{Neal A. Miller\altaffilmark{1}} 
\email{nmiller@pha.jhu.edu}

\author{Ann E. Hornschemeier\altaffilmark{1,2}}

\author{Bahram Mobasher\altaffilmark{3}}

\altaffiltext{1}{Department of Physics and Astronomy, Johns Hopkins University, 3400 N. Charles Street, Baltimore, MD 21218}
\altaffiltext{2}{Laboratory for X-Ray Astrophysics, NASA Goddard Space Flight Center, Code 662.0, Greenbelt, MD 20771}
\altaffiltext{3}{Department of Physics \& Astronomy, University of California, Riverside, CA 92521}

\begin{abstract} 

We present deep 1.4~GHz Very Large Array (VLA) radio continuum observations of two $\sim$half square degree fields in the Coma cluster of galaxies. The two fields, ``Coma 1'' and ``Coma 3,'' correspond to the cluster core and southwest infall region and were selected on account of abundant pre-existing multiwavelength data. In their most sensitive regions the radio data reach 22 $\mu$Jy rms per 4\farcs4 beam, sufficient to detect (at 5$\sigma$) Coma member galaxies with $L_{1.4~GHz} = 1.3 \times 10^{20}$ W Hz$^{-1}$. The full catalog of radio detections is presented herein and consists of 1030 sources detected at $\geq 5\sigma$, 628 of which are within the combined Coma 1 and Coma 3 area. We also provide optical identifications of the radio sources using data from the Sloan Digital Sky Survey (SDSS). The depth of the radio observations allows us to detect AGN in cluster elliptical galaxies with $M_r < -20.5$ (AB magnitudes), including radio detections for all cluster ellipticals with $M_r < -21.8$. At fainter optical magnitudes ($-20.5 < M_r \lesssim -19$) the radio sources are associated with star-forming galaxies with star formation rates as low as 0.1 \myr.

\end{abstract}

\section{Introduction}\label{sec-intro}

The Coma cluster of galaxies, Abell 1656 \citep[][hereafter referred to as Coma]{aco1989}, has been among the most well-studied galaxy clusters \citep[e.g., see the review by][]{biviano1998}. At $z=0.0231$ \citep{struble1999} it is nearby, yet significantly richer than Virgo and more accessible than Perseus (A426), which lies near the Galactic Plane. Studies of Coma have shaped our understanding of galaxy evolution in the cluster environment. In assessing the frequency of blue galaxies in intermediate redshift clusters, Coma was one of several nearby clusters used as a benchmark for establishing the ``Butcher-Oemler Effect'' \citep{butcher1984}. Although the fraction of blue galaxies decreases with decreasing redshift and therefore the activity level of Coma is presumably low, studies have still revealed galaxy evolution in progress. \citet{caldwell1993,caldwell1996,caldwell1999} and \citet{caldwell1997,caldwell1998} have studied images and spectroscopy of Coma galaxies to reveal populations of starbursts and post-starburst galaxies, and related these findings to substructures within Coma and hence connected the star-formation history of galaxies to cluster environment. In particular, Coma member starburst and post-starburst galaxies are preferentially located in the regions around NGC~4839 which appears to be a large infalling group \citep{colless1996,neumann2003} although others have argued for this system being seen shortly after passage through the core of Coma \citep{burns1994}. Similar studies have contrasted Coma with higher redshift clusters; at higher redshift more luminous galaxies are observed to constitute the starburst and post-starburst populations \citep[e.g.,][]{dressler1999}, while sources in Coma with similar spectroscopic signatures are fainter dwarf galaxies \citep{poggianti2004}. This suggests that ``downsizing'' -- the shift in star-formation activity to lower mass systems -- is occurring, and that some of the Butcher-Oemler effect is the result of applying a fixed absolute magnitude cutoff to both low and high redshift systems.

Recently, Coma has been studied to greater depth using the William Herschel Telescope (WHT). \citet{komiyama2002} presented optical imaging in five $32\farcm5 \times 50\farcm8$ fields, with \citet{mobasher2001} using the imaging to select targets for spectroscopy for the ``core'' and ``outskirt'' fields: ``Coma 1'' centered at 12$^{\mbox{{\scriptsize h}}}$59$^{\mbox{{\scriptsize m}}}$45\fs2 $+$27$^\circ$57\arcmin53\farcs1, and ``Coma 3'' at 12$^{\mbox{{\scriptsize h}}}$57$^{\mbox{{\scriptsize m}}}$28\fs5 $+$27$^\circ$08\arcmin08\farcs5 and having NGC~4839 near its northern edge. The southwest corner of the Coma 1 field overlaps slightly with the northeast corner of the Coma 3 field, meaning that combined they define a contiguous total area of about 0.92 square degrees (refer to Figure \ref{fig-rmsconts}). These photometric and spectroscopic data were used to construct the cluster optical luminosity function \citep{mobasher2003}, study stellar population ages and metallicities \citep{poggianti2001a,poggianti2001b}, investigate the effect of environment on galaxy properties and compare the findings to higher redshift clusters \citep{carter2002,poggianti2004}, and examine cluster dynamics for both giant galaxies and dwarfs \citep{edwards2002}.

Surprisingly, the best wide area 1.4~GHz radio continuum survey of Coma to date is the Faint Images of the Radio Sky at Twenty centimeters survey \citep[FIRST;][]{first}. With a limiting sensitivity of 1 mJy and 5\arcsec{} resolution, the FIRST data are capable of detecting Coma member sources with $L_{1.4~GHz} \gtrsim 1.2 \times 10^{21}$ \whz. This amounts to radio luminosities about half that of the Milky Way and greater \citep{condon1992}, or in terms of a star formation rate (SFR) about 0.7 \myr \citep[assuming a Salpeter IMF from 0.1 to 100 M$_\odot$, using the relationship of][]{yun2001}. While this is a useful limit for studies of active galaxy populations in clusters \citep[e.g.,][identify an average of $\sim$10 radio-detected galaxies per cluster down to this limit and within each cluster's Abell radius]{miller2002}, the proximity of Coma makes it relatively easy to study hitherto unexplored fainter radio galaxy populations. Deeper 1.4~GHz observations in the vicinity of NGC~4869, the prototypical head-tail radio galaxy at the core of Coma, are presented in \citet{feretti1995}. These data reach an rms sensitivity of about 30 $\mu$Jy at about the same resolution as FIRST, and have the added benefit of full polarization information. They are of very limited areal coverage (a few square arcminutes), however, on account of bandwidth smearing. Similarly, the excellent {\scshape H~i} observations of spiral galaxies in Coma have narrow total bandwidths and hence relatively poor radio continuum sensitivity \citep[][the rms for 1.4~GHz continuum of these data is about 0.2 mJy]{bravoalfaro2000,bravoalfaro2001}. These data show evidence for cluster environmental effects in the form of ram pressure stripping, including asymmetric {\scshape H~i} morphologies and greater gas deficiencies for galaxies near the core of the cluster.

There is also a wealth of new multiwavelength data available for Coma. X-ray satellites have imaged both the core and large-scale regions of the cluster \citep[e.g.,][]{white1993,briel2001,vikhlinin2001,neumann2003,finoguenov2004}, and much deeper observations aimed at the Coma 3 region have been performed with {\it Chandra} \citep{hornschemeier2006} and {\it XMM} (Hornschemeier et al., in preparation). Fewer prior data exist in the ultraviolet (UV), although Coma was observed by the FOCA balloon-borne imaging telescope \citep{donas1995}. However, new UV data from {\it GALEX} including a deep observation centered on Coma 3 are now available and indicate a faint-end slope to the UV LF that is steeper than comparable field UV LFs \citep{hammer2007}. Although more traditionally associated with star formation, the UV data from {\it GALEX} detect numerous elliptical galaxies, even down to less luminous, dwarf-like objects. The reason for this is emission from extreme horizontal branch stars, known as the ``UV upturn'' in their spectra \citep{greggio1990}. In addition to the aforementioned WHT optical programs, a new {\it Hubble Space Telescope} Treasury program \citep{carter2007} obtained two-color imaging of portions of Coma using the {\it Advanced Camera for Surveys}. The Sloan Digital Sky Survey \citep[SDSS;][]{york2000} includes optical imaging and spectroscopy for Coma in its fifth and sixth data releases \citep[DR5 and DR6;][]{adelman2007,sdssdr6}. In the infrared (IR), {\it Spitzer} is producing interesting results on the cluster. Near-IR observations with IRAC (3.6 -- 8.0 $\mu$m) are reported in \citet{jenkins2007} for the Coma 1 and Coma 3 regions, and also suggest a steep faint-end slope for the near-IR LF. These LF results are indicative of an excess of dwarf galaxies relative to optically-selected samples. At longer IR wavelengths, \citet{bai2006} constructed 24 $\mu$m and 70 $\mu$m LFs using {\it Spitzer} MIPS and found their overall shape to be consistent with field LFs. They did, however, find evidence for environmental effects in the form of different slopes for the core and infall regions.

To complement these deep and wide area multi-waveband datasets, we used the National Radio Astronomy Observatory (NRAO)\footnote{The National Radio Astronomy Observatory is a facility of the National Science Foundation operated under cooperative agreement by Associated Universities, Inc.} Very Large Array (VLA) to obtain deep 1.4~GHz images of the core (Coma 1) and outskirts (Coma 3) of the Coma cluster. Our observations reach 22 $\mu$Jy, or $L_{1.4~GHz} \approx 1.3 \times 10^{20}$ \whz{} for galaxies at the distance of Coma. For Coma members powered by star formation, this equates to SFR $\gtrsim 0.08$ \myr. \citet{hopkins2002} find the median SFR of blue compact dwarf galaxies as determined by 1.4~GHz emission to be 0.3 \myr, implying that our VLA survey should be sufficiently sensitive to provide radio detections for some of the brighter dwarfs, and useful upper limits for non-detections. The radio source catalog derived from our VLA observations has been correlated with the SDSS optical data to explore the radio galaxy population of Coma, and by extension numerous radio sources lying behind the cluster. 

In this paper, we present the observations, data reduction, and resulting source catalogs. A related paper investigates the characteristics of the radio-detected Coma member galaxies and derives the cluster radio luminosity function \citep{rlf}. For all distance-dependent calculations, we assume a redshift of $z=0.0231$ corresponding to a distance modulus of 35 using the standard WMAP cosmology \citep{spergel2007}. This is consistent with distances to Coma derived using a host of other techniques \citep[e.g.,][]{gregg1997,hjorth1997,kavelaars2000,tully2000}.

\section{Radio Data}

\subsection{Observations}

The VLA observations were performed over five days in June 2006 as program code AM868. On each of the five days the scheduled time was centered on the transit of Coma. The VLA was in its B, or second largest, configuration. In order to obtain nearly-uniform sensitivity across the $\sim30\arcmin{} \times \sim50\arcmin$ areas corresponding to Coma 1 and Coma 3, mosaics of individual pointings were made. These amounted to 11 separate pointings for each field, arranged in a hexagonal grid with 12\farcm5 between adjacent pointings (see Table \ref{tbl-points} and Figure \ref{fig-rmsconts}). This grid spacing insures that all regions of the Coma 1 and Coma 3 fields with the exception of small areas in six of the corners lie within the half-power response of more than one pointing. Observations of 3C286 for primary flux density calibration and 1310+323 for bandpass and phase calibration meant that each of the 22 total pointings received just under one hour of total integration. To minimize the time spent slewing while still obtaining good (u,v) coverage, we devoted individual days to Coma 1 and Coma 3 (2 days each) and reobserved all 22 pointings on the fifth day. On days when a single field was targetted we observed each of its 11 pointings twice for about 11.5 minutes each time, with the order shuffled on the other date that field was observed. Thus, in total each pointing received five individual $\sim11.5$ minute observations at five different hour angles, resulting in excellent total (u,v) coverage as demonstrated in Figure \ref{fig-uvcover} for the centermost pointing within the Coma 1 field.

The correlator was operated in the standard multichannel continuum mode used for deep, wide-field 1.4~GHz observations. This corresponds to seven 3.125~MHz channels in each of two bands (called ``intermediate frequencies'' or IFs) centered at 1.365 and 1.435~GHz, respectively, with each IF receiving both left and right circular polarizations (hence 28 total channels). The smaller channels alleviate bandwidth smearing while still providing a fairly large total bandwidth and thus increasing the sensitivity. At the time of the observations the VLA had just begun its transition to the Expanded VLA (``EVLA''), with individual VLA antennas being retrofitted with EVLA receivers and electronics. Depending on the date of the observations either three or four EVLA anntennas were used in the observations on a shared-risk basis, along with 22 standard VLA antennas. The EVLA antennas were operated in the same manner as the standard VLA antennas (e.g., same observed frequency channels), so the benefits of including them were simply the increased number of array elements.

\subsection{Data Reduction}

The data were reduced using the Astronomical Image Processing Software (AIPS) package, version 31DEC06. The flux density scale was tied to 3C286, which was bootstrapped to the phase calibrator 1310+323. This latter calibrator was observed approximately every 45 minutes and also was used for calibrating the bandpasses. Some flagging of data was necessary, usually for the EVLA antennas. For example, one EVLA antenna (the one most recently brought into service) provided useable data for only one IF and polarization. The others were reliable with the exception of short periods of time usually at the conclusion of each day's observations.

Once the data were calibrated, the initial steps toward producing final images were made by pointing ID and observation date. Thus, there were 66 separate (u,v) datasets with which to work, corresponding to 11 pointings each for the first four observing dates and all 22 pointings on the last (refer to Table \ref{tbl-points}). The data for each pointing were imaged using the task IMAGR with a strong taper to produce an image 2\degr{} on a side, onto which was overlaid a ``flys-eye'' pattern of 37 slightly overlapping 512\arcsec$\times$512\arcsec{} square facets. The apparent positions of bright sources outside the coverage of the flys-eye were noted, and their true positions were taken from the FIRST catalog \citep{first}. There were usually around ten such bright sources located outside the central flys-eye pattern. In subsequent imaging, all 37 facets of the flys-eye and all the bright sources outside of it were imaged separately using the 3D option of IMAGR. Additionally, a few of the brightest sources within the flys-eye coverage were also given their own facets for imaging.

Each of the 66 (u,v) datasets were then subjected to iterative rounds of imaging and self calibration. After initially imaging the $\sim50$ facets for a specific pointing, the imaged facets were inspected and sources were boxed for subsequent imaging rounds. The imaged facets were used as inputs for self calibration and additional editting of the (u,v) data. The end product was 66 well-behaved (u,v) datasets, and the three separate days of data per pointing were then combined to form the semi-final 22 (u,v) datasets (i.e., one per mosaic pointing). Each of these were again subjected to imaging and self calibration, with the final images per pointing having sensitivities ranging from 39 to 45 $\mu$Jy per 4\farcs4 beam. The observational strategy ensured nearly identical (u,v) coverage for each pointing, resulting in circular 4\farcs4 restoring beams for all pointings.

We also performed quick tests of the relative performance of the retrofitted EVLA antennas. These were achieved by imaging the data for a given pointing both with and without the EVLA antennas included. The rms noise of the images which included the EVLA antennas was better by the expected amount under the assumption that they were identical to the standard VLA antennas. We note that the main benefit of the retrofitted antennas at 1.4~GHz is their performance over wider bandwidths, which will be available for use when the EVLA's new correlator comes online. For the current project, we proceed treating the retrofitted EVLA antennas no differently than the regular VLA ones.

The final image was created by assembling the mosaic of these 22 individual pointing images using the task FLATN. For this mosaic, two $4096 \times 4096$ images with 1\arcsec{} pixels were created, centered on Coma 1 and Coma 3 (12$^{\mbox{{\scriptsize h}}}$59$^{\mbox{{\scriptsize m}}}$45\fs2 $+$27$^\circ$57\arcmin53\farcs1  and 12$^{\mbox{{\scriptsize h}}}$57$^{\mbox{{\scriptsize m}}}$28\fs5 $+$27$^\circ$08\arcmin08\farcs5, respectively). Only data out to the 30\% power point of the VLA 1.4~GHz primary beam were used, which corresponds to the central 19 facets of each flys-eye. Data for each pointing were variance weighted according to the VLA primary beam power pattern. The rms sensitivity of the final image is fairly uniform, as shown in Figure \ref{fig-rmsconts} which depicts contours of constant rms as evaluated over 5\arcmin{} square boxes centered on each location. The majority of the formal Coma 1 and Coma 3 areas have rms below 29 $\mu$Jy (Figure \ref{fig-rmsarea}), which correponds to a 5$\sigma$ detection of a Coma cluster galaxy forming stars at 0.1 \myr{} \citep[assuming the conversion of][]{yun2001}. The deepest regions within Coma 1 have an rms sensitivity of 22 $\mu$Jy, while those in Coma 3 are slightly less sensitive (Figure \ref{fig-rmsarea}). This is largely on account of the sidelobe response of the bright extended radio galaxies in the center of Coma 1 (NGC4874 and NGC4869).

\subsection{Source Catalog}

With varying sensitivity across the full mosaic image, we opted to create signal-to-noise ($S/N$) maps for source detection. We first calculated the local rms for each pixel in the final mosaic images, evaluated over a 5\arcmin{} box using the task RMSD (refer to Figure \ref{fig-rmsconts}; smaller-sized boxes are overly effected by real extended sources such as NGC4869). We then divided each mosaic image by its corresponding rms image to obtain the $S/N$ maps. Source detection was then accomplished via the task SAD, which was directed to identify sources with peaks greater than 4.5$\sigma$ and fit them with Gaussians. SAD was run on six separate portions of the full mosaic images to insure that sources were not missed due to the task's limitations for the maximum number of sources it will evaluate in a given run. We also had SAD create residual images, the purpose of which will be described shortly. The procedure is largely the same as that described in \citet{biggs2006}.

The SAD catalogs then served as input for creating the final catalog. Each source in the SAD catalogs was examined in the mosaic images (not the $S/N$ image) using the task JMFIT, and its position, peak flux density, integral flux density and error, whether the source was resolved, and whether it was near the edge of the mosaic (within $\sim$2\farcm5, where the local rms measurement would be based on fewer resolution elements) were recorded. Repeated astrometric measurements with the VLA at 1.4~GHz indicate a positional accuracy of around 0\farcs1 for strong unresolved point sources. The fainter sources in our survey have positional errors on the order of 0\farcs5 \citep[Equation 1 of][]{condon1998}. Because JMFIT uses the noise measurement made over the full image when reporting flux density errors, we used the measured value at the source position in the RMSD map as the error in the peak flux density measurement. Similarly, we updated the integral flux density errors by scaling the JMFIT-reported values by the ratio of the local noise to the full mosaic area noise. Assessing whether a source is resolved is not entirely straightforward due to the mosaic nature of the images. Although greatly reduced by the smaller channels, bandwidth smearing does smear sources radially -- and hence in different orientations for each pointing's contribution to a given source. Thus, an intrinsically unresolved source might appear to be marginally resolved on account of bandwidth smearing arising in the individual pointings that contributed to that source within the mosaic. We call a source resolved if the minimum size of its reported major axis from JMFIT was non-zero, however we note that there will exist false positives in the catalog. These are usually brighter sources, and can often be spotted through comparison of their peak and integral flux densities, which will be equivalent within their respective errors for unresolved sources.

The residual images were then inspected to identify sources either missed or poorly fit by SAD. This was greatly simplified by the procedure of having run source detection on the $S/N$ map, since contours at specific values of $\sigma$ could be plotted and thereby render the identification of such sources easy. Those sources missed by SAD were added to the final catalog after repeating the procedure outlined in the preceding paragraph. Because both SAD and JMFIT rely on Gaussian fitting, plotted $\sigma$ contours on the residual map also ease identification of sources for which fitting an individual Gaussian was problematic. In such cases, JMFIT was used to deblend sources which were better described as multiple overlapping Gaussians. Finally, for the more irregularly shaped extended sources (e.g., NGC4869) the task TVSTAT was used to perform aperture photometry. The catalog positions for these sources correspond to the location of their peak flux density.

The resulting radio source catalog is presented in Table \ref{tbl-radiocat}. There are over 1200 sources listed, although this figure includes lower significance detections and multiple entries for some sources which appear to be associated with a single extragalactic object (e.g., one entry for each lobe of a radio double). Not all sources that were fit by multiple components by SAD consist of multiple entries in Table \ref{tbl-radiocat}, however. In cases such as NGC4874 where we have prior knowledge of the source and its origin in a single radio galaxy we have included only a single entry in Table \ref{tbl-radiocat}.  We have attempted to provide guidance for the possible multi-component sources by noting where our visual inspection of the images suggested a relationship between separate entries in the radio source catalog. This may be found in the ``Comments'' field of Table \ref{tbl-radiocat}. The table includes sources down to $S/N \ge 4.5\sigma$, where $S/N$ is based on the peak flux density and error from the SAD run on the $S/N$ map. Although 5$\sigma$ is traditionally used in radio surveys, we include sources with slightly lower significance. There are two reasons for this: First, they still have a very high chance of being non-spurious, especially in cases where they coincide with detected sources at other wavelengths. Second, diffuse sources with peak flux densities less than five times the local noise might have integral flux densities greater than five times their integral flux density errors. Restricting our acceptance criteria to sources with $S/N \ge 5\sigma$ and within the 0.92 square degree contiguous area formed by the Coma 1 and Coma 3 fields yields 628 radio sources.

\section{Matching to Optical Data}

The Sloan Digital Sky Survey \citep[SDSS;][]{york2000} included the photometry for Coma in its Data Release 5 \citep[DR5;][]{adelman2007}. The SDSS observes the sky in drift-scan mode, using a novel multi-CCD camera \citep{gunn1998} that simultaneously collects data in five filters ($u$, $g$, $r$, $i$, and $z$). The median seeing at $r$ is 1\farcs4, and the 95\% completeness limit for point source detection is 22.2 mag in this filter \citep{adelman2007}. The magnitudes are on the AB system \citep{oke1983} and have uncertainties in the 0.02 -- 0.03 mag range \citep{ivezic2004}. Catalog data are available online ({\ttfamily http://www.sdss.org/dr5/}) and are described in the Early Data Release paper \citep{stoughton2002}. For correlation with the radio data, we used the SDSS DR5 catalog of all objects within large boxes that extended $\gtrsim 10\arcmin$ past each side of the formal Coma 1 and Coma 3 regions. These included both photometrically-classified stars and galaxies.

\begin{figure}
\figurenum{4}
\epsscale{0.9}
\plotone{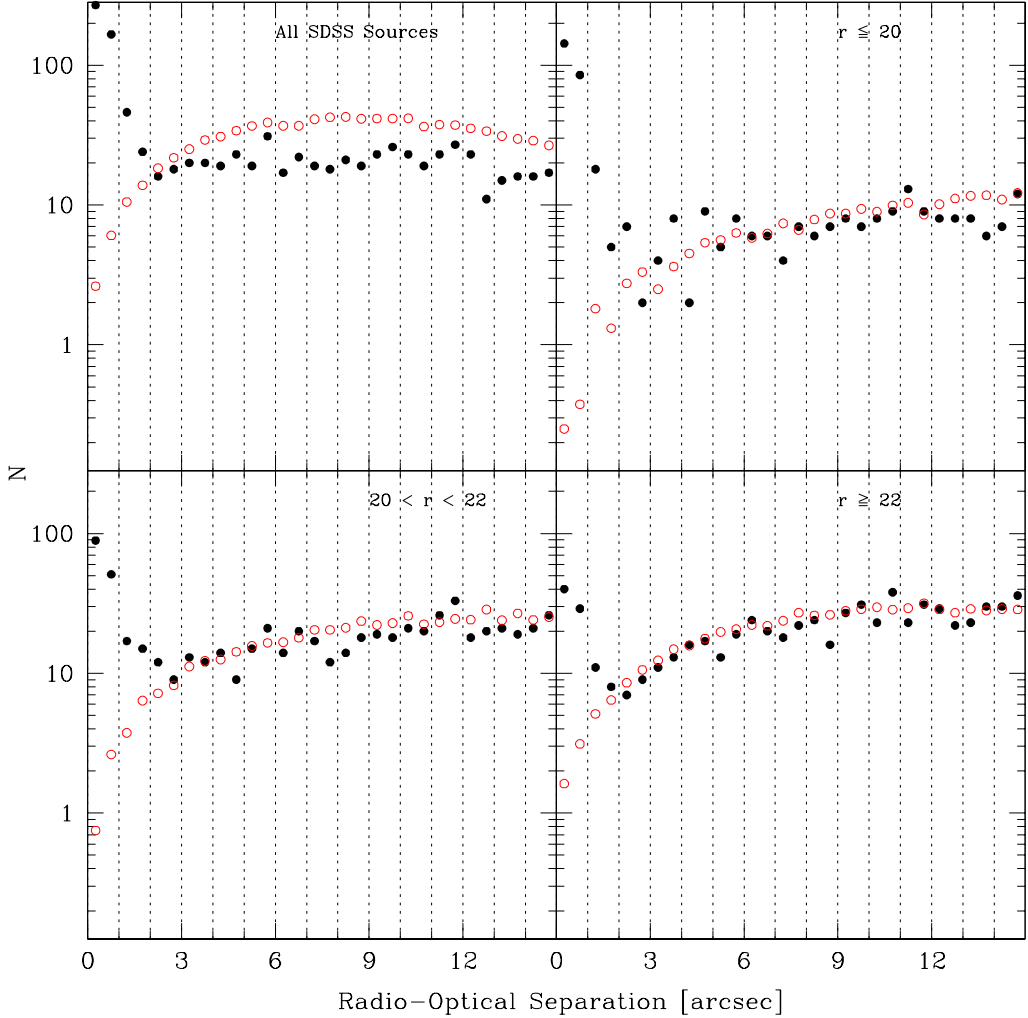}
\caption{Number of radio-optical pairs as a function of separation. The results for the actual radio catalog are plotted as filled circles, while the frequency of chance superpositions as evaluated by applying shifts to the radio catalog are shown as open circles. Top left: for all SDSS sources; top right: for SDSS sources with $r \leq 20$; bottom left: for SDSS sources with $20 < r < 22$; bottom right: for SDSS sources with $r \geq 22$.\label{fig-offsets}}
\end{figure}

We adopted a matching radius of 2\arcsec{} for sources with $r\leq22$ and 1\arcsec{} for fainter sources. These figures were determined based on direct estimation using the data. First, Table \ref{tbl-radiocat} was correlated with the SDSS optical catalog, and the SDSS object nearest to each radio source was recorded. To evaluate whether a given radio-optical pair represented a real physical association, we repeated this correlation a number of times after having applied arbitrary fixed offsets to the radio catalog, thus arriving at estimates of the probability of chance superpositions derived directly from the data. The resulting distribution of radio-optical separations is shown in the top-left panel of Figure \ref{fig-offsets}. The large peak in the distribution of the unshifted data with small radio-optical separations is obvious, and the vast majority of such objects are real associations of radio emission with an optical source. At larger separations the two distributions are approximately the same, indicating that sources at such separations are consistent with expectations for random superposition of a radio source with an optical source. The false match rate appears higher than the actual data for most of the separations plotted in the top left panel of Figure \ref{fig-offsets} because the number of sources used in generating the shifted catalogs is conserved and the fraction of radio sources with real optical counterparts is high. In other words, if we were to remove the real counterparts and renormalize the unshifted data to have the same number of sources as the shifted data, the two curves would be more coincident. The two curves cross at a separation near 2\arcsec, indicating that radio sources with optical counterparts closer than this value have a high probability of being physically associated with one another while radio sources with more distant optical counterparts are likely false associations. The density of sources in the radio and optical catalogs is sufficiently high that at the larger plotted separations the number of matches declines.

The use of the 1\arcsec{} search radius for fainter objects was determined in the same manner. Figure \ref{fig-offsets} also includes the real and offset catalog results for three separate magnitude ranges, $r \leq 20$ (top right), $20 < r < 22$ (bottom left), and $r \geq 22$ (bottom right). It can be seen that for each of the $r \leq 20$ and $20 < r < 22$ samples the distributions of the real data and the shifted data cross somewhere just below 3\arcsec. However, for the fainter sources ($r \geq 22$) the crossover is at separations less than 2\arcsec. Figure \ref{fig-offsets} also provides a direct way to estimate the confidence with which a given radio-optical association may be judged. For example, a pair with a 1\arcsec{} separation would have about a 1 in 40 chance of being a false association if the optical source had $r \leq 20$, about a 1 in 10 chance if $20 < r < 22$, and about a 1 in 5 chance if $r \geq 22$. A comparison of the three panels also shows the expected result that the optically brighter galaxies are more likely to be associated with a radio source. 

The $S/N$ map that was used in the generation of the radio source catalog was then overlaid on the $r$-band images from the SDSS. The entire area of the radio mosaic was thus inspected to identify apparent matches of radio emission with optical sources, guided by an input listing of all SDSS sources with radio counterparts from Table \ref{tbl-radiocat} within 3\arcsec. As indicated, we generally accepted sources with radio-optical separations up to 2\arcsec{} for brighter objects ($r < 22$, or $M_r = -13$ in Coma) but were more restrictive for fainter objects and used 1\arcsec{} separation as the cut. Some sources with separations smaller than these values were rejected for a variety of reasons. In some cases, multiple SDSS counterparts of a radio source existed and the radio emission was ascribed to the object that was deemed the more likely counterpart, with such a determination usually based on giving preference to the nearer counterpart with an SDSS object type of galaxy. Seventeen sources with separations greater than 2\arcsec{} were also accepted in cases where the radio emission was extended and its morphology was consistent with expectations based on the optical source. Examples include spiral galaxies with extended radio emission matching their optical disks (e.g., NGC4911) and elliptical galaxies hosting ``FR1'' and ``FR2'' type radio sources (e.g., NGC4869, NGC4874). All of these exceptions are depicted in Figure \ref{fig-extrads}. Finally, the visual inspection found four radio doubles where the midpoint of the two radio positions was within 1\arcsec{} of the apparent optical counterpart. 

The full radio-optical source catalog is presented in Table \ref{tbl-radopt}, and consists of 499 optical counterparts to radio sources from Table \ref{tbl-radiocat}. For completeness, we present the rejected optical sources within 3\arcsec{} of a radio source in Table \ref{tbl-rejects} along with brief explanations for their exclusion from Table \ref{tbl-radopt}. The table includes summary data for the SDSS optical counterpart, including its photometric classification (galaxy or star) and $ugriz$ ``model'' magnitudes. These are based on a common aperture defined by the more likely of a deVaucouleurs or pure exponential profile fit to the data at $r$ band, and are the recommended magnitudes for color studies of galaxies \citep[e.g., see][]{adelman2007}.

\section{Discussion and Summary}

Unless otherwise noted, in the ensuing discussion we restrict our analysis to the contiguous $\sim0.9$ square degree area defined by the combined Coma 1 and Coma 3 fields (see Section \ref{sec-intro} and Figure \ref{fig-rmsconts}) and refer to this generically as the ``survey area.'' 

About half of the radio sources within the survey area have optical counterparts in the SDSS. Of the 628 radio sources with $S/N \geq 5\sigma$, 312 have optical counterparts from the SDSS catalog and an additional three optical counterparts to radio sources were found in our visual inspection of the radio data overlaid onto the SDSS $r$-band images (refer to Table \ref{tbl-radopt}; the optical sources not found within the SDSS catalog were presumably excluded due to their location near very bright objects such as diffraction spikes around saturated stars). A further 22 radio sources detected at less than 5$\sigma$ were also identified with SDSS optical counterparts within the survey area.

\begin{figure}
\figurenum{6}
\plotone{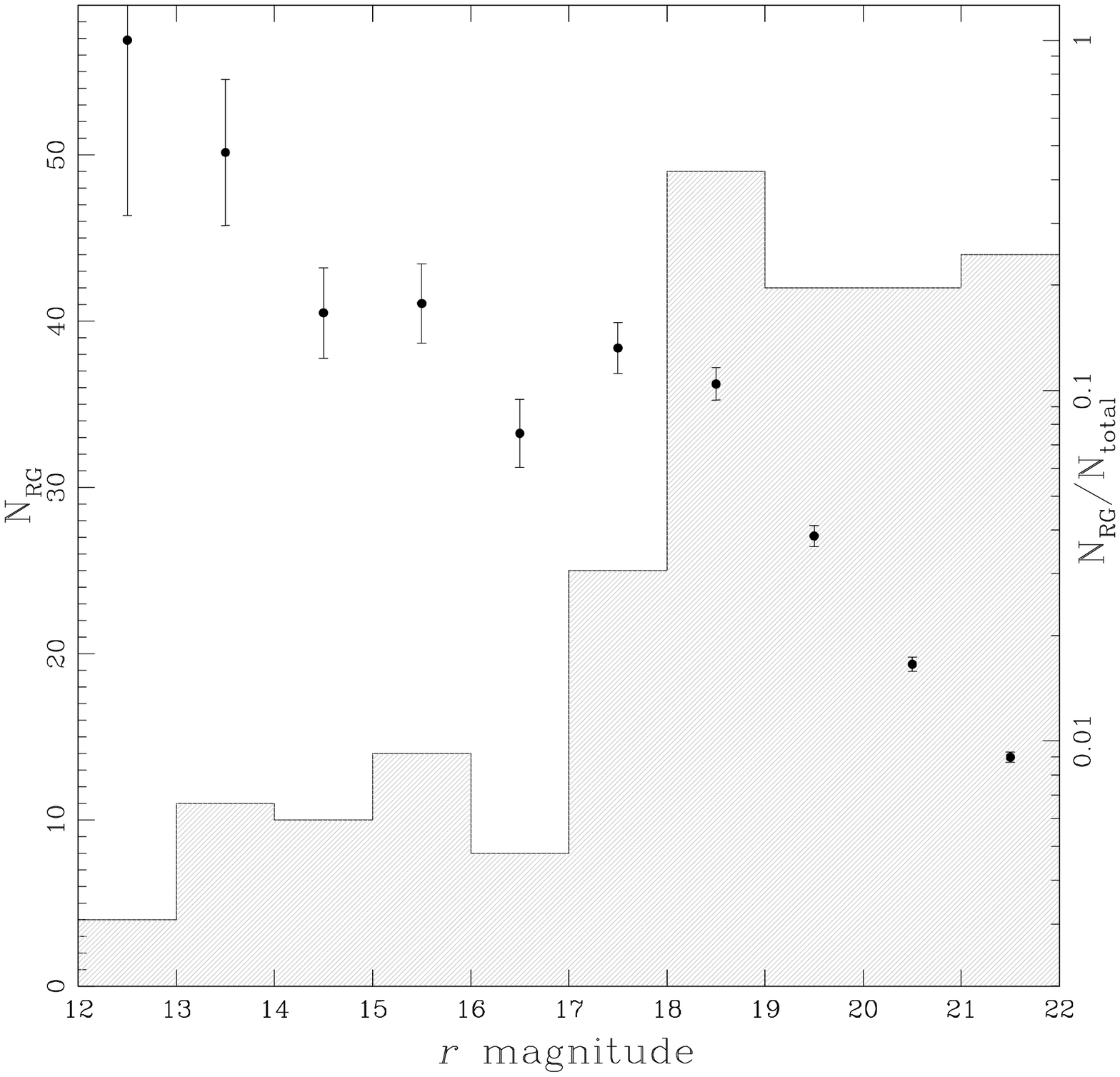}
\caption{Histogram of $r$ magnitudes for SDSS galaxies with radio counterparts for sources within the Coma 1 and Coma 3 area (left vertical axis indicates number of radio galaxies, $N_{RG}$), and fraction of total SDSS galaxies ($N_{RG}$/$N_{tot}$) detected by radio observations within that magnitude bin (right vertical axis).\label{fig-magdist}}
\end{figure}

Conversely, the fraction of optical sources ($N_{tot}$) detected by the radio observations ($N_{RG}$) is shown in Figure \ref{fig-magdist}. Specifically, we use the SDSS objects with photometric classification as galaxies\footnote{Six bright objects ($r<12$) were removed from the optical catalog as inspection of the images indicated they were saturated stars mistakenly classified as galaxies.} and investigate the fraction of sources with $\geq5\sigma$ radio detections as a function of $r$ magnitude. At the brighter magnitudes ($r\leq14$, or $M_r \leq -21$ for Coma members), our survey detects over half of the galaxies (15/28, including all six objects with $r\leq13.04$). Velocity measurements for the radio detections, including many new measurements for fainter galaxies determined from an MMT campaign using Hectospec (Marzke et al. in preparation), are presented in \citet{rlf} where we derive and discuss the radio LF. We note here that NED contains measurements for most of the galaxies in the present discussion (those with $r \lesssim 17.5$) since Coma has been the target of numerous galaxy spectroscopic surveys, including the SDSS \citep{sdssdr6}. Assuming a redshift range 4,000 \kms{}$\leq cz \leq$ 10,000 \kms{} for Coma \citep[e.g.,][]{colless1996}, all 15 of the radio detections with $r\leq14$ are found to be Coma members. The radio detection fraction remains in about the 10\% - 20\% range for $14 < r \leq 19$ before dropping. At brighter optical magnitudes the radio sources are almost exclusively Coma members, while at the fainter magnitudes they are predominantly background galaxies. By $r\approx21$ the radio-detected fraction falls to about 1\%.

\begin{figure}
\figurenum{7}
\plotone{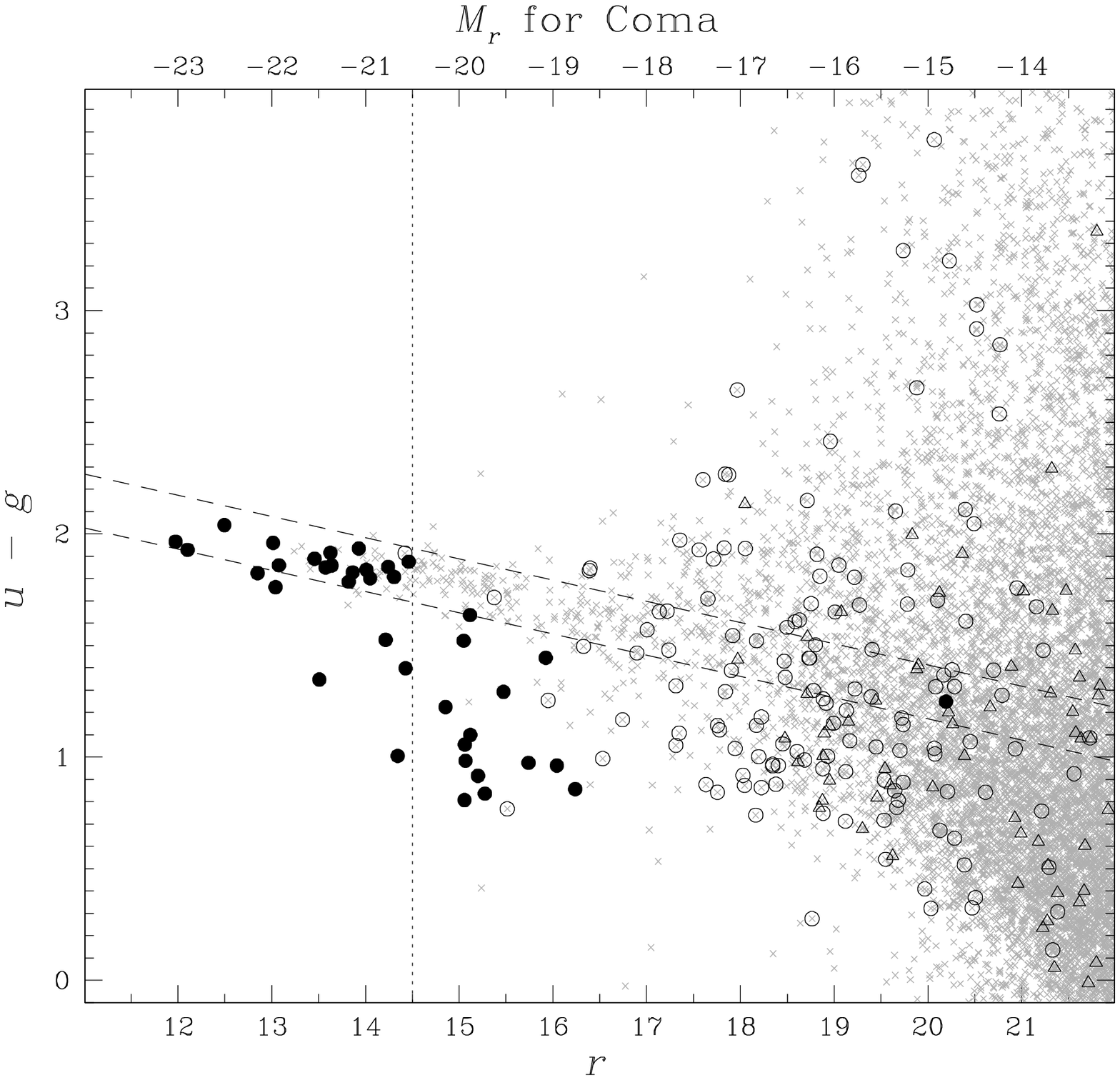}
\caption{Color magnitude diagram for all SDSS galaxies within the Coma 1 and Coma 3 regions. Non-radio detected galaxies are shown as small grey crosses, while radio-detected galaxies are indicated by larger symbols: filled circles are spectroscopically-confirmed Coma members (4,000 \kms $\leq cz \leq$ 10,000 \kms), open circles are spectroscopically-confirmed background galaxies (with the exception of the blue object at $r = 15.5$, which is the planetary nebula PG1257+279), and open triangles are sources for which no spectroscopy is available (see \citet{rlf} for details on spectroscopy of Coma radio-detected galaxies). The long dashed lines are the $\pm 2\sigma$ boundaries to the fitted Coma red sequence, while the vertical dotted line shows the approximate limit for radio detection of red sequence members.\label{fig-cmd}}
%REF next paper??
\end{figure}

The types of the radio-detected sources are elucidated by their optical colors. Figure \ref{fig-cmd} shows the optical color-magnitude diagram, with radio detections indicated by filled circles for cluster members, open circles for non-members, and open triangles for sources without velocity information. The red sequence is very clearly delineated, and we fitted it for galaxies with $r \leq 17$ using the procedure of \citet{lopezcruz2004}. Note that this fitting is strictly based on photometric data, so some galaxies that lie on the red sequence will actually be non-cluster members whose colors happen to be consistent with the Coma red sequence. It can be seen from Figure \ref{fig-cmd} that the radio detection fraction for red sequence galaxies rises at $r = 14.5$ ($M_r = -20.5$ for Coma). Setting the red sequence range at $\pm2\sigma$ about the fitted relation of $(u - g) = 3.192 - 0.095r$, 18 of 43 red sequence galaxies with $r < 14.5$  are radio detections. The detection rate increases with optical luminosity, as found in numerous other studies \citep[e.g.,][]{ledlow1996}. All five red sequence galaxies having $r < 13.2$ ($M_r < -21.8$) are detected by the radio data (these include four of the five brightest cluster galaxies noted above). Conversely, only two of the 135 red sequence galaxies with $14.5 \leq r \leq 17$ ($-18 \geq M_r \geq -20.5$) are radio detections. Based on velocities from the literature, one of the 18 radio detections among the brighter galaxies is a background source \citep[J12581169+2708117 with $cz=11889$ \kms;][]{sdssdr6} along with one of the two radio-detected fainter galaxies \citep[J12581684+2710267 with $cz=18179$ \kms;][]{colless1996}. This means that the faintest radio-detected Coma member on the red sequence has $M_r = -19.9$, with the possible exception of J12591536+2746052. \citet{biviano1995} report a velocity of 6315 \kms{} for this galaxy, although they indicate this determination is of intermediate quality as it is based on few spectral features. If true, it would have $M_r = -14.8$ and $L_{1.4}=2.1\times10^{20}$ \whz, making it an extreme outlier about four magnitudes fainter than the next faintest Coma radio-detected member.

The detection fraction is similar for galaxies bluer than the red sequence. Of the 66 galaxies with $r \leq 17$ and ($u - g$) colors at least $2\sigma$ bluer than the fitted red sequence, 25 are radio detections. Most of the detections have $r \leq 16$, where 17 radio detections are found among 34 total galaxies. Again, relying on published velocity measurements from NED and the SDSS we find that 6 of the 25 radio-detected galaxies are background galaxies (2 of 17 for the $r \leq 16$ detections). The faintest radio-detected Coma member with a blue color has $M_r = -18.8$. Presumably, the radio emission from these blue Coma galaxies has its origins in star formation. This is largely confirmed via further analysis of photometric data and optical spectroscopy in \citet{rlf}, where only one blue Coma galaxy showed spectroscopic evidence for an AGN in the form of emission line diagnostics typical of a Seyfert galaxy (ARK395). Its blue colors (including UV) extending across its disk suggest that star formation also contributes to the net radio emission from this galaxy. Based on their respective radio and optical morphologies, some of the blue Coma galaxies appear to be undergoing ram pressure stripping. For example, Figure \ref{fig-extrads} includes IC4040 which has radio contours compressed to its northwest side and extended to the southeast -- consistent with the {\scshape H~i} deficit and morphology reported in \citet{bravoalfaro2000}. This disturbed morphology produces an offset between the fitted radio position and the nucleus of the galaxy and hence its inclusion in Figure \ref{fig-extrads}. Although it lies outside the Coma 3 footprint, KUG1255+283 (also shown in Figure \ref{fig-extrads}) has a similar radio morphology, in this case with its radio emission extended away from the galaxy in a direction roughly opposite that of the cluster center. Other examples of known star-forming galaxies with radio contours suggestive of ram pressure interactions include Markarian 57 and KUG1258+279A with radio-optical positional offsets of 1\farcs3 and 1\farcs0, respectively. Again, in each case the radio contours are asymmetric with respect to the galaxy nucleus and extended in a direction roughly pointing away from the center of the cluster. Both are included in the \citet{bravoalfaro2000} work, although Markarian 57 is actually {\scshape H~i} rich. KUG1258+279A is {\scshape H~i} deficient, although only weakly detected and hence its {\scshape H~i} morphology is only suggestive of ram pressure interaction \citep{bravoalfaro2001}.

% 300x300 images on narya:
%  Mrk 57 is PL1 for Coma3_R.FIN_R.1, base contour at 30uJy, sqrt transfer
%    usual 2 3 5 8...89 contours
%  KUG1258+279A base at 31uJy, PL3 for COMA1_R.FIN_R.1
% Redo on elessar? MRK57 12:58:37.2 +27:10:35
%                  KUG   13:00:33.7 +27:38:16
% MRK57 is very HI rich in Bravo-alfaro; HI extended to N... opposite 
% radio continuum contours?
% also note C3C-910, in figure of large offsets, is second faintest
%  object with cluster velocity. Looks good - very blue edge-on spiral,
%  very weak radio that is extended and matches disk.

\begin{figure}
\figurenum{3}
\epsscale{0.9}
\plotone{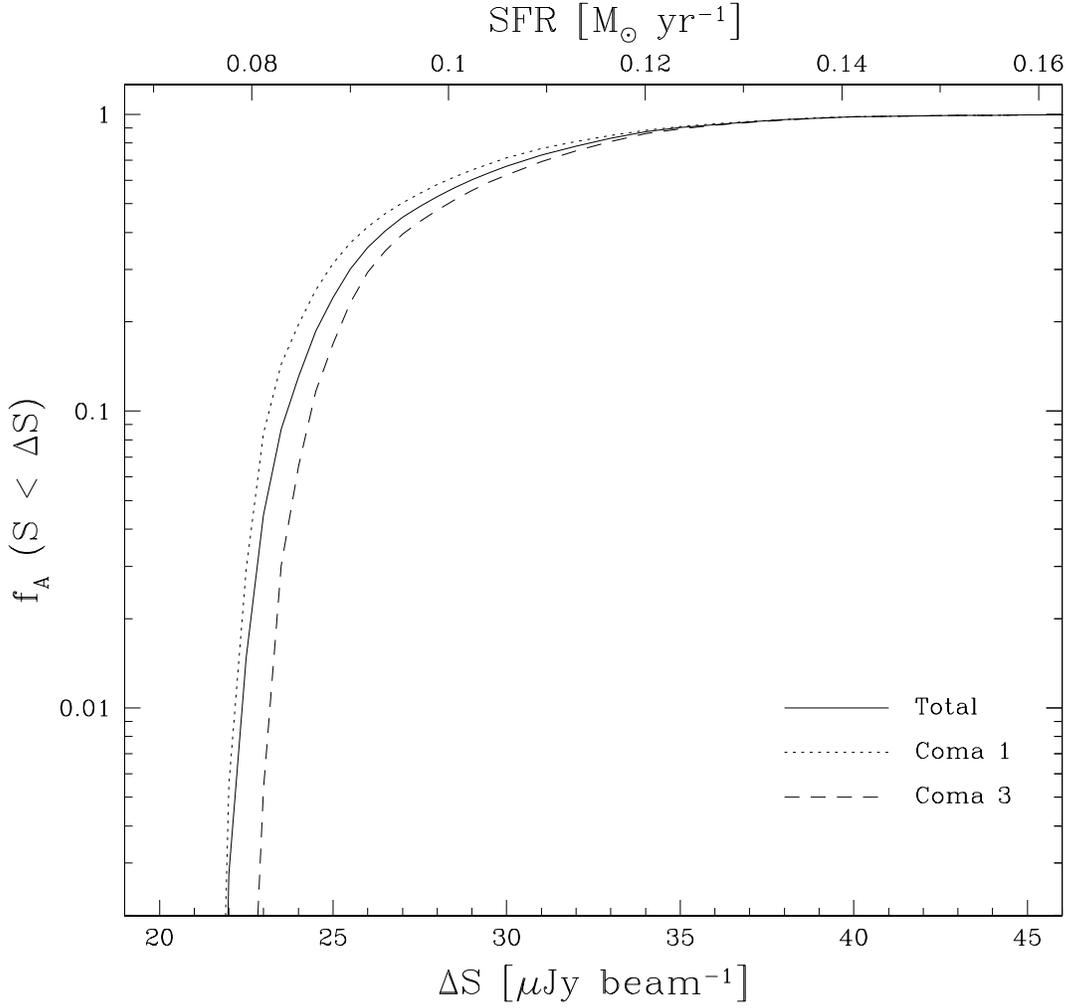}
\caption{Fractional area covered at a given sensitivity limit for Coma 1 (dotted line), Coma 3 (dashed line), and combined (solid line). The top axis is the associated star formation rate for a source detected at 5$\sigma$, determined using the 1.4~GHz star formation rate calibration from \citet{yun2001}.\label{fig-rmsarea}}
\end{figure}

Although this survey specifically addresses the Coma cluster, its combination of sensitivity and areal coverage are comparable to some published deep fields. The total area covered within the combined Coma 1 and Coma 3 footprint is about 0.9 square degrees, with a typical rms sensitivity of 28 $\mu$Jy per 4\farcs4 beam (refer to Figures \ref{fig-rmsconts} and \ref{fig-rmsarea}). The VLA-VIRMOS Deep Field (VVDS) includes radio observations with an rms sensitivity of 17 $\mu$Jy per 6\arcsec{} beam across a 1 square degree region \citep{bondi2003}, and the Subaru/XMM-Newton Deep Field (SXDF) covers 0.8 square degrees with 20 $\mu$Jy sensitivity at about 5\arcsec{} by 4\arcsec{} resolution \citep[note that these figures are specifically for the subset of the total area surveyed having local rms sensitivity below 20 $\mu$Jy;][]{simpson2006}. Along with the excellent multiwavelength data available for Coma (refer to Section \ref{sec-intro}) including optical spectroscopy of galaxies down to fairly faint magnitudes ($r \sim 21$; Marzke et al., in preparation), the multi-waveband Coma data will also be of value to deep field science.
 
% Bondi et al. has 1054 sources in their 5sig 1 deg sq field
% Simpson et al. has 505 with flux over 100 uJy in that 0.8 deg sq field
% Hopkins et al. two regions: wide area, about 4.6 square deg at 60 uJy, and
%  deeper down to 12 uJy. Beam is 6 x 12.

\begin{figure}
\figurenum{8}
\plotone{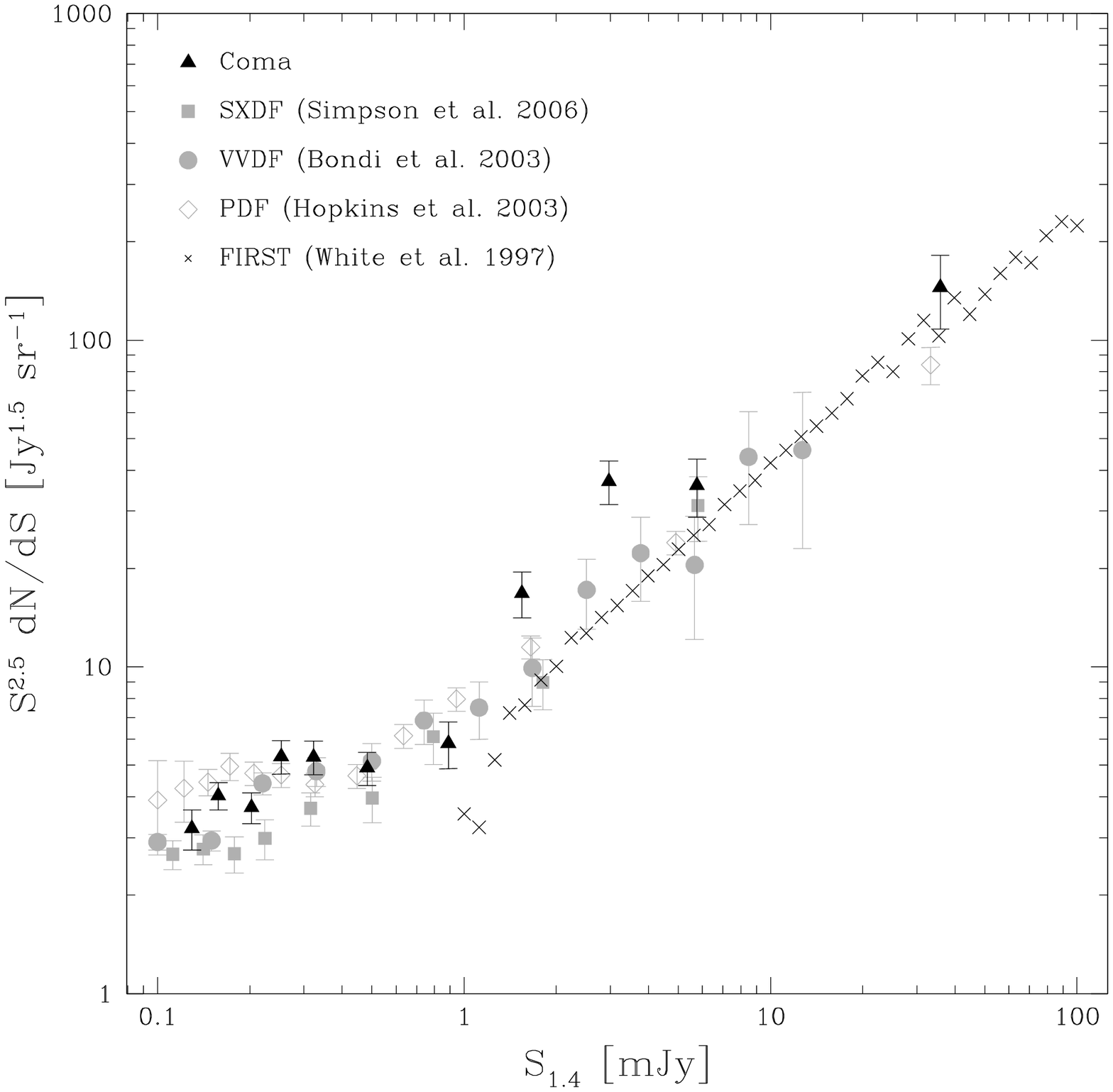}
\caption{Source counts for Coma data (filled black triangles) along with those derived from similar surveys.\label{fig-src}}
\end{figure}

Figure \ref{fig-src} shows the source counts for our survey, and the source count data are provided in Table \ref{tbl-src}. We have restricted this analysis to the 628 sources within the Coma 1 and Coma 3 coverage area that were detected with a peak flux density greater than five times the local noise. In addition to source counts from the VVDS and SXDF, we also plot those of the Phoenix Deep Field \citep[PDF;][]{hopkins2003} and the FIRST survey. The wider area of these surveys improves the numbers and hence error bars for source counts at higher flux densities. We note that the source counts for the Coma data have only been corrected for the areal coverage in each bin (i.e., sources in the fainter bins are detectable only in the deeper regions of the survey area), thus ignoring effects such as Eddington bias and resolved sources with integral flux densities sufficient for inclusion but peak flux densities failing to reach the 5$\sigma$ cutoff. Even so, the source counts are consistent with these other studies, showing the usual decline in normalized counts between 100 mJy and 1 mJy, followed by leveling below 1 mJy. A slight excess of sources is seen in the $\sim1$ to $\sim5$ mJy region, and it is tempting to attribute this to Coma member galaxies since these flux densities correspond to luminosities of a few times $10^{21}$ \whz{} for galaxies at the distance of Coma. While Coma galaxies certainly do contribute to the source counts between $\sim1$ to $\sim5$ mJy, the excess over the FIRST survey source counts in this region is much greater than can be explained by the presence of the Coma cluster. For the source counts at approximately 1.5 mJy and 3 mJy in Figure \ref{fig-src}, the difference between the number of sources in the Coma survey area and that expected based on the FIRST source counts amounts to roughly 20 to 25 sources per each bin. The matching to SDSS optical data and optical spectroscopic databases indicates that only nine Coma members have radio flux densities falling in these bins (three for the 1.5 mJy bin and six for the 3 mJy bin). The comprehensive optical spectroscopic coverage for radio-detected galaxies \citep[see][]{rlf} insures that we are not missing Coma galaxies down to very faint optical magnitudes ($M_r \lesssim -15$).

\acknowledgments
NAM gratefully acknowledges the support of a Jansky Fellowship of the NRAO, held during the period when the majority of this work was performed. The authors thank Dave Carter and the members of the HST/ACS Coma Treasury Program.

Funding for the Sloan Digital Sky Survey (SDSS) has been provided by the Alfred P. Sloan Foundation, the Participating Institutions, the National Aeronautics and Space Administration, the National Science Foundation, the U.S. Department of Energy, the Japanese Monbukagakusho, and the Max Planck Society. The SDSS Web site is http://www.sdss.org/.

The SDSS is managed by the Astrophysical Research Consortium (ARC) for the Participating Institutions. The Participating Institutions are The University of Chicago, Fermilab, the Institute for Advanced Study, the Japan Participation Group, The Johns Hopkins University, the Korean Scientist Group, Los Alamos National Laboratory, the Max-Planck-Institute for Astronomy (MPIA), the Max-Planck-Institute for Astrophysics (MPA), New Mexico State University, University of Pittsburgh, University of Portsmouth, Princeton University, the United States Naval Observatory, and the University of Washington.

This research has made use of the NASA/IPAC Extragalactic Database (NED) which is operated by the Jet Propulsion Laboratory, California Institute of Technology, under contract with the National Aeronautics and Space Administration.

\clearpage

\begin{figure}
\figurenum{1}
\epsscale{0.9}
%\plotone{f1.ps}
\caption{Contours of constant rms sensitivity for final mosaic image. Rectangular areas for Coma 1 (upper left) and Coma 3 (lower right) are indicated, along with crosses marking the locations of the 22 VLA pointing centers. The innermost contour for each Coma 1 and Coma 3 is at 25 $\mu$Jy, with successive contours at 30, 35, 40, 45, 50, and 55 $\mu$Jy.\label{fig-rmsconts}}
\end{figure}

\begin{figure}
\figurenum{2}
\epsscale{0.85}
%\rotatebox{270}{\plotone{f2.ps}}
\caption{The (u,v) coverage for the central pointing in the Coma 1 mosaic (Coma1\_F in Table \ref{tbl-points}). Every 51$^{st}$ record is plotted.\label{fig-uvcover}}
\end{figure}

\begin{figure}
\figurenum{5}
%\plotone{f5a.ps}
\caption{Radio galaxies with separations over 2\arcsec. Each square image is 2\arcmin{} by 2\arcmin, with radio contours at 3, 5, 8, 13, ... times the local noise. The top right galaxy is IC3913, and the middle left is NGC4839. The galaxy at the bottom left is KUG1255+283, and appears to be undergoing ram pressure stripping.\label{fig-extrads}}
\end{figure}

\begin{figure}
\figurenum{5}
%\plotone{f5b.ps}
\caption{continued. The galaxy in the middle right panel is NGC4874, and the bottom panel is NGC4869.}
\end{figure}

\begin{figure}
\figurenum{5}
%\plotone{f5c.ps}
\caption{continued. The galaxy in the middle right panel is IC4040, which appears to be undergoing ram pressure stripping. The bottom left galaxy is NGC4911.}
\end{figure}

\clearpage

\begin{deluxetable}{c l l l}
\tablecolumns{4}
\tablecaption{Observation Summary\label{tbl-points}}
\tablewidth{0pt}
\tablehead{
\colhead{Pointing ID} & \colhead{RA(J2000)} & \colhead{Dec(J2000)} & 
\colhead{Dates Observed}}
\startdata
Coma1\_A & 13 00 34.40 & +28 16 38.1 & 2006 Jun 22, 2006 Jun 24, 2006 Jun 25 \\
Coma1\_B & 12 58 56.00 & +28 16 38.1 & 2006 Jun 22, 2006 Jun 24, 2006 Jun 25 \\
Coma1\_C & 12 59 45.20 & +28 10 23.1 & 2006 Jun 22, 2006 Jun 24, 2006 Jun 25 \\
Coma1\_D & 13 00 34.30 & +28 04 08.1 & 2006 Jun 22, 2006 Jun 24, 2006 Jun 25 \\
Coma1\_E & 12 58 56.10 & +28 04 08.1 & 2006 Jun 22, 2006 Jun 24, 2006 Jun 25 \\
Coma1\_F & 12 59 45.20 & +27 57 53.1 & 2006 Jun 22, 2006 Jun 24, 2006 Jun 25 \\
Coma1\_G & 13 00 34.20 & +27 51 38.1 & 2006 Jun 22, 2006 Jun 24, 2006 Jun 25 \\
Coma1\_H & 12 58 56.20 & +27 51 38.1 & 2006 Jun 22, 2006 Jun 24, 2006 Jun 25 \\
Coma1\_I & 12 59 45.20 & +27 45 23.1 & 2006 Jun 22, 2006 Jun 24, 2006 Jun 25 \\
Coma1\_J & 13 00 34.10 & +27 39 08.1 & 2006 Jun 22, 2006 Jun 24, 2006 Jun 25 \\
Coma1\_K & 12 58 56.30 & +27 39 08.1 & 2006 Jun 22, 2006 Jun 24, 2006 Jun 25 \\
Coma3\_A & 12 58 17.30 & +27 26 53.5 & 2006 Jun 18, 2006 Jun 23, 2006 Jun 25 \\
Coma3\_B & 12 56 39.70 & +27 26 53.5 & 2006 Jun 18, 2006 Jun 23, 2006 Jun 25 \\
Coma3\_C & 12 57 28.50 & +27 20 38.5 & 2006 Jun 18, 2006 Jun 23, 2006 Jun 25 \\
Coma3\_D & 12 58 17.20 & +27 14 23.5 & 2006 Jun 18, 2006 Jun 23, 2006 Jun 25 \\
Coma3\_E & 12 56 39.80 & +27 14 23.5 & 2006 Jun 18, 2006 Jun 23, 2006 Jun 25 \\
Coma3\_F & 12 57 28.50 & +27 08 08.5 & 2006 Jun 18, 2006 Jun 23, 2006 Jun 25 \\
Coma3\_G & 12 58 17.10 & +27 01 53.5 & 2006 Jun 18, 2006 Jun 23, 2006 Jun 25 \\
Coma3\_H & 12 56 39.90 & +27 01 53.5 & 2006 Jun 18, 2006 Jun 23, 2006 Jun 25 \\
Coma3\_I & 12 57 28.50 & +26 55 38.5 & 2006 Jun 18, 2006 Jun 23, 2006 Jun 25 \\
Coma3\_J & 12 58 17.00 & +26 49 23.5 & 2006 Jun 18, 2006 Jun 23, 2006 Jun 25 \\
Coma3\_K & 12 56 40.00 & +26 49 23.5 & 2006 Jun 18, 2006 Jun 23, 2006 Jun 25 \\
\enddata

\tablecomments{The pointings for each field are arranged in three north-south columns where A-D-G-J form the first column, C-F-I from the middle column, and B-E-H-K form the last column. The ``F'' pointings thus correspond to the formal Coma 1 and Coma 3 field centers from \citet{komiyama2002}.}

\end{deluxetable}

\begin{deluxetable}{c l l r r r r r r r l}
\tablecolumns{11}
\tablecaption{Radio Source Catalog\label{tbl-radiocat}}
\tabletypesize{\footnotesize}
\tablewidth{0pt}
\tablehead{
\colhead{ID} & \colhead{RA(J2000)} & \colhead{Dec(J2000)} & 
\colhead{$S/N$} & \colhead{$S_{peak}$} & \colhead{rms} & 
\colhead{$S_{int}$} & \colhead{$\Delta S_{int}$} & 
\colhead{Res?} & \colhead{Edge?} & \colhead{Comment}}
\startdata
C3A-001 & 12 55 11.22 & 27 27 22.4 &   7.9 &   921 &  115 &   1880 &  313 & 1 & 1 & \nodata \\
C3A-002 & 12 55 12.38 & 27 27 39.8 &   4.8 &   556 &  114 &    570 &  201 & 1 & 1 & \nodata \\
C3B-917 & 12 55 12.46 & 26 59 39.6 &   4.6 &   551 &  119 &   1082 &  316 & 1 & 1 & \nodata \\
C3B-001 & 12 55 12.57 & 26 58 58.4 &  12.6 &  1516 &  120 &   3562 &  443 & 1\tablenotemark{a} & 1 & \nodata \\
C3B-003 & 12 55 13.30 & 26 59 48.4 &   8.5 &  1007 &  118 &   2236 &  358 & 1\tablenotemark{a} & 1 & \nodata \\
\enddata

\tablenotetext{a}{Extended source with integral flux density and error determined in irregularly-shaped aperture using TVSTAT.}

%\tablenotetext{b}{Although the peak $S/N < 5$ in the $S/N$ map, the fitted value from JMFIT is above a 5$\sigma$ detection.}
 
%\tablenotetext{c}{Although the peak $S/N \ge 5$ in the $S/N$ map, the fitted value from JMFIT is below a 5$\sigma$ detection.}

\tablecomments{ONLY A PORTION OF THE TABLE IS SHOWN HERE. Column descriptions: (1) unique source ID that we found useful in our analyis; (2) \& (3) right ascension and declination (J2000) of radio source fitted peak; (4) $S/N$ of fitted peak as determined by source identification algorithm (SAD) on $S/N$ map (see text). Sources with $4.94 < S/N < 5.0$ are listed as $S/N = 4.9$ to clarify that they are not formally 5$\sigma$ detections.; (5) peak flux density in $\mu$Jy; (6) local rms as evaluated in a 5\arcmin{} square box centered on the source position (this is also the error associated with the peak flux density measurement); (7) \& (8) integral flux density and its associated error, each in $\mu$Jy; (9) binary code indicating whether the source was resolved ($=1$) or unresolved ($=0$); (10) binary code indicting whether the source was within $\sim$2\farcm5 of the edge of the radio mosaic ($=1$); and (11) comments.}

\end{deluxetable}

\begin{deluxetable}{r r r r r r r r r r l}
\tablecolumns{11}
\tablecaption{Optical Counterparts to Radio Sources\label{tbl-radopt}}
\tabletypesize{\footnotesize}
\tablewidth{0pt}
\tablehead{
\colhead{RA} & \colhead{Dec} & \colhead{Sep} &
\colhead{$u$} & \colhead{$g$} & \colhead{$r$} & \colhead{$i$} & \colhead{$z$} & 
\colhead{Class} & \colhead{Radio ID} & \colhead{Comments}}
\startdata
12 55 17.18 & 26 45 48.2 & 0.39 & 24.956 & 24.918 & 23.207 & 22.029 & 21.345 & 6 & C3C-002 & \nodata \\
12 55 20.64 & 27 14 09.2 & 0.72 & 22.006 & 21.768 & 21.761 & 21.148 & 21.238 & 6 & C3B-006 & C3B-005, C3B-009? \\
12 55 23.33 & 27 02 13.9 & 0.48 & 23.073 & 21.895 & 20.626 & 20.015 & 19.275 & 3 & C3B-011 & \nodata \\
12 55 23.90 & 26 56 40.9 & 0.71 & 24.052 & 21.926 & 20.222 & 19.377 & 18.987 & 3 & C3C-003 & \nodata \\
12 55 24.10 & 26 57 37.4 & 0.41 & 23.894 & 24.526 & 24.003 & 22.211 & 21.630 & 6 & C3B-012 & \nodata \\
\enddata

\tablecomments{ONLY A PORTION OF THE TABLE IS SHOWN HERE. Column descriptions: (1) Right ascension of SDSS optical counterpart (J2000); (2) Declination of SDSS optical counterpart (J2000); (3) Separation between radio position and SDSS optical counterpart, in arcsec; (4 - 8) SDSS $ugriz$ model magnitudes; (9) SDSS photometric class of optical counterpart, where 3 = galaxy and 6 = star; (10) Radio ID, from Table \ref{tbl-radiocat}; (11) Comments, where:\\
a = Radio emission is resolved, providing a likely explanation for the somewhat large radio-optical separation.\\
b = The morphology of the extended radio emission is consistent with that of the optical. In the case of spiral galaxies this indicates that the radio emission traces the galaxy disk, whereas for elliptical galaxies the center of the galaxy appears to be where the radio emission originates.\\
c = Source appears to be the host of a radio double. The indicated separation is for distance from optical source to midpoint of the radio counterparts listed in columns 10 and 11.\\
d = Source separation is greater than 3\arcsec. See Figure \ref{fig-extrads}.\\
e = Optical position for bright clump within larger galaxy.\\
f = Galaxy appears to be one member of a galaxy pair.\\
g = Radio emission is unresolved, but lies within the optical extent of the galaxy.\\
h = Optical source not in SDSS catalog due to proximity to bright source, usually the diffraction spike of a saturated star. The position represents the coordinates of the peak of the object as measured directly in the SDSS $r$ band image.\\
}

\end{deluxetable}

\begin{deluxetable}{r r r r r r r r r r l}
\tablecolumns{11}
\tablecaption{Rejected Optical Counterparts to Radio Sources\label{tbl-rejects}}
\tabletypesize{\footnotesize}
\tablewidth{0pt}
\tablehead{
\colhead{RA(J2000)} & \colhead{Dec(J2000)} & \colhead{Sep} &
\colhead{$u$} & \colhead{$g$} & \colhead{$r$} & \colhead{$i$} & \colhead{$z$} & 
\colhead{Class} & \colhead{Radio ID} & \colhead{Comments}}
\startdata
12 55 33.15 & 26 41 14.3 & 1.04 & 23.394 & 22.900 & 21.996 & 21.205 & 20.494 & 3 & C3C-007 & a \\
12 55 45.55 & 26 50 05.3 & 1.27 & 26.666 & 23.608 & 21.443 & 20.295 & 19.620 & 3 & C3C-010 & a \\
12 55 47.06 & 27 27 50.8 & 2.32 & 21.903 & 20.786 & 19.543 & 19.016 & 18.544 & 3 & C3A-013 & \nodata \\
12 55 49.34 & 27 33 30.6 & 2.49 & 23.872 & 21.690 & 20.717 & 20.523 & 20.023 & 3 & C3A-015 & b \\
12 55 58.32 & 27 21 23.0 & 2.45 & 24.888 & 24.555 & 22.777 & 21.511 & 20.637 & 6 & C3A-020 & b \\
\enddata

\tablecomments{ONLY A PORTION OF THE TABLE IS SHOWN HERE. Column descriptions: (1) Right ascension of rejected SDSS optical counterpart (J2000); (2) Declination of rejected SDSS optical counterpart (J2000); (3) Separation between radio position and SDSS optical counterpart, in arcsec; (4 - 8) SDSS $ugriz$ ``cmodel'' magnitudes; (9) SDSS photometric class of optical counterpart, where 3 = galaxy and 6 = star; (10) Radio ID, from Table \ref{tbl-radiocat}; (11) Comments, where:\\
a = Although the radio-optical separation is acceptable for the $r$ magnitude of the potential counterpart, visual inspection of the overlaid radio contours strongly suggested the sources are unassociated.\\
b = The radio source was assigned to a galaxy with a smaller separation (see Table \ref{tbl-radopt}).\\ 
c = Same comment as ``a'' plus the radio source is detected at less than $5\sigma$.\\
d = The radio morphology is that of a radio double source, and this optical counterpart appears to be a chance superposition of a galaxy with one of the lobes. The optical counterpart for the radio emission would be expected to lie in between the lobes, and not coincident with one.\\
e = Visually appears offset. The radio emission also appears to be a lobe of a separate FR2 type source, and hence not associated with this galaxy.\\
f = This radio source was assigned to a brighter galaxy with radio-optical separation greater than 3\arcsec. It may be seen in Figure \ref{fig-extrads}.\\
}

\end{deluxetable}

%source counts: usual cols are 1: S (range of bin); 2: Save (ave flux of
% gals in this bin); 3: N (number in bin); 4: Correction factor; 5: 
% S^2.5 dN/dS (Jy^1.5 sr^-1)
\begin{deluxetable}{r r r r r}
\tablecolumns{5}
\tablecaption{Source Counts within Coma 1 and Coma 3\label{tbl-src}}
\tablewidth{0pt}
\tablehead{
\colhead{$S$} & \colhead{$<S>$} & \colhead{$N$} & \colhead{Factor} & 
\colhead{$S^{2.5}dN/dS$}\\
\colhead{[mJy]} & \colhead{[mJy]} & \colhead{} & \colhead{} & 
\colhead{[Jy$^{1.5}$sr$^{-1}$]}}
\startdata
0.110 - \phn0.140 & 0.129 &  50 & 2.81 &  3.20$\pm$\phn0.45\\
0.140 - \phn0.180 & 0.158 & 109 & 1.32 &  4.04$\pm$\phn0.39\\
0.180 - \phn0.230 & 0.202 &  87 & 1.02 &  3.71$\pm$\phn0.40\\
0.230 - \phn0.280 & 0.253 &  73 & 1.00 &  5.32$\pm$\phn0.62\\
0.280 - \phn0.370 & 0.323 &  71 & 1.00 &  5.30$\pm$\phn0.63\\
0.370 - \phn0.650 & 0.483 &  75 & 1.00 &  4.91$\pm$\phn0.57\\
0.650 - \phn1.200 & 0.889 &  38 & 1.00 &  5.83$\pm$\phn0.95\\
1.200 - \phn2.000 & 1.546 &  40 & 1.00 & 16.8\phn$\pm$\phn2.7\phn\\
2.000 - \phn4.000 & 2.969 &  43 & 1.00 & 37.0\phn$\pm$\phn5.6\phn\\
4.000 - 10.000    & 5.757 &  24 & 1.00 & 36.0\phn$\pm$\phn7.3\phn\\
10.000 - 99.999   & 35.783 & 15 & 1.00 & 145.\phn\phn$\pm$37.\phn\phn\\
\enddata

\end{deluxetable}


\begin{thebibliography}{dummy}

\bibitem[Abell, Corwin, \& Olowin(1989)]{aco1989} Abell, G.O., Corwin, H.G., \& Olowin, R.P. 1989, \apjs, 70, 1

\bibitem[Adelman-McCarthy et al.(2007)]{adelman2007} Adelman-McCarthy, J.K. et al. 2007, \apjs, 172, 634 (SDSS DR5)

\bibitem[Adelman-McCarthy et al.(2008)]{sdssdr6} Adelman-McCarthy, J.K. et al. 2008, \apjs, 175, 297 (SDSS DR6)

\bibitem[Bai et al.(2006)]{bai2006} Bai, L., Rieke, M.J., Hinz, J.L., Kelly, D.M., \& Blaylock, M. 2006, \apj, 639, 827

\bibitem[Becker, White, \& Helfand(1995)]{first} Becker, R.H., White, R.L., \& Helfand, D.J. 1995, \apj, 450, 559 (FIRST)

\bibitem[Biggs \& Ivison(2006)]{biggs2006} Biggs, A.D., \& Ivison, R.J. 2006, \mnras, 371, 963

\bibitem[Biviano(1998)]{biviano1998} Biviano, A. 1998, in ``Untangling Coma Berenices: A New Vision of an Old Cluster,'' eds. Mazure, A., Casoli, F., Durret, F., Gerbal, D., (Singapore:World Scientific)

\bibitem[Biviano et al.(1995)]{biviano1995} Biviano, A., Durret, F., Gerbal, D., Le~Fevre, O., Lobo, C., Mazure, A., \& Slezak, E. 1995, \aaps, 111, 265

\bibitem[Blanton et al.(2003)]{blanton2003} Blanton, M.R., Brinkmann, J., Csabai, I., Doi, Mamoru, Eisenstein, D., Fukugita, M., Gunn, J.E., Hogg, D.W., \& Schlegel, D.J. 2003, \aj, 125, 2348 ({\ttfamily kcorrect})

\bibitem[Bondi et al.(2003)]{bondi2003} Bondi, M., et al. 2003, \aap, 403, 857

\bibitem[Bravo-Alfaro et al.(2000)]{bravoalfaro2000} Bravo-Alfaro, H., Cayatte, V, van Gorkum, J.H., \& Balkowski, C. 2000, \aj, 119, 580

\bibitem[Bravo-Alfaro et al.(2001)]{bravoalfaro2001} Bravo-Alfaro, H., Cayatte, V, van Gorkum, J.H., \& Balkowski, C. 2001, \aap, 379, 347

\bibitem[Briel et al.(2001)]{briel2001} Briel, U.G., Henry, J.P., Lumb, D.H., Arnaud, M., Neumann, D., Aghanim, N., Gastaud, R., Mittaz, J.P.D., Sasseen, T.P., \& Vestrand, W.T. 2001, \aap, 365, L60

\bibitem[Burns et al.(1994)]{burns1994} Burns, J.O., Roettiger, K., Ledlow, M., \& Klypin, A. 1994, \apj, 427, L87

\bibitem[Butcher \& Oemler(1984)]{butcher1984} Butcher, H., \& Oemler, A. 1984, \apj, 285, 426

\bibitem[Caldwell et al.(1993)]{caldwell1993} Caldwell, N., Rose, J.A., Sharples, R.M., Ellis, R.S., \& Bower, R.G. 1993, \aj, 106, 473

\bibitem[Caldwell et al.(1996)]{caldwell1996} Caldwell, N., Rose, J.A., Franx, M., \& Leonardi, A. 1996, \aj, 111, 78

\bibitem[Caldwell \& Rose(1997)]{caldwell1997} Caldwell, N., \& Rose, J.A. 1997, \aj, 113, 492

\bibitem[Caldwell \& Rose(1998)]{caldwell1998} Caldwell, N., \& Rose, J.A. 1998, \aj, 115, 1423

\bibitem[Caldwell et al.(1999)]{caldwell1999} Caldwell, N., Rose, J.A., \& Dendy, K. 1999, \aj, 117, 140

\bibitem[Carter et al.(2002)]{carter2002} Carter, D., Mobasher, B., Bridges, T.J., Poggianti, B.M., Komiyama, Y., Kashikawa, N., Doi, M., Iye, M., Okamura, S., Sekiguchi, M., Shimasaku, K., Yagi, M., \& Yasuda, N. 2002, \apj, 567, 772

\bibitem[Carter et al.(2008)]{carter2007} Carter, D. et al. 2008, \apjs, in press

\bibitem[Colless \& Dunn(1996)]{colless1996} Colless, M., \& Dunn, A.M. 1996, \apj, 458, 435

\bibitem[Condon(1992)]{condon1992} Condon, J.J. 1992, \araa, 30, 575

\bibitem[Condon et al.(1998)]{condon1998} Condon, J.J., Cotton, W.D., Greisen, E.W., Yin, Q.F., Perley, R.A., Taylor, G.B., \& Broderick, J.J. 1998, \aj, 115, 1693

\bibitem[Donas et al.(1995)]{donas1995} Donas, J., Milliard, B., \& Laget, M. 1995, \aap, 303, 661

\bibitem[Dressler et al.(1999)]{dressler1999} Dressler, A., Smail, I., Poggianti, B.M., Butcher, H., Couch, W., Ellis, R., \& Oemler, A. 1999, \apjs, 122, 51

\bibitem[Edwards et al.(2002)]{edwards2002} Edwards, S.A., Colless, M., Bridges, T.J., Carter, D., Mobasher, B., \& Poggianti, B.M. 2002, \apj, 567, 178

\bibitem[Feretti et al.(1995)]{feretti1995} Feretti, L., Dallacasa, D., Giovannini, G., \& Tagliani, A. 1995, \aap, 302, 680

\bibitem[Finoguenov et al.(2004)]{finoguenov2004} Finoguenov, A., Briel, U.G., Henry, J.P., Gavazzi, G., Iglesias-Paramo, J., \& Boselli, A. 2004, \aap, 419, 47

\bibitem[Gregg(1997)]{gregg1997} Gregg, M.D. 1997, New Astronomy, 1, 363

\bibitem[Greggio \& Renzini(1990)]{greggio1990} Greggio, L., \& Renzini, A. 1990, \apj, 364, 35

\bibitem[Gunn et al.(1998)]{gunn1998} Gunn, J.E. et al. 1998, \aj, 116, 3040

\bibitem[Hammer et al.(2008)]{hammer2007} Hammer, D. M., Hornschemeier, A.E., Salim, S., Jenkins, L., Mobasher, B., Miller, N., Ferguson, H., \& Heckman, T. 2008, \aj, in submission

\bibitem[Hjorth \& Tanvir(1997)]{hjorth1997} Hjorth, J., \& Tanvir, N.R. 1997, \apj, 482, 68

\bibitem[Hopkins et al.(2003)]{hopkins2003} Hopkins, A.M., Afonso, J., Chan, B., Cram, L.E., Georgakakis, A., \& Mobasher, B. 2003, \aj, 125, 465

\bibitem[Hopkins et al.(2002)]{hopkins2002} Hopkins, A.M., Schulte-Ladbeck, R.E., \& Drozdovsky, I.O. 2002, \aj, 124, 862

\bibitem[Hornschemeier et al.(2006)]{hornschemeier2006} Hornschemeier, A.E., Mobasher, B., Alexander, D.M., Bauer, F.E., Bautz, M.W., Hammer, D., Poggianti, B.M. 2006, \apj, 643, 144

\bibitem[Ivezi\'c et al.(2004)]{ivezic2004} Ivezi\'c et al. 2004, Astronomiche Nachrichten, 325, 583

\bibitem[Jenkins et al.(2007)]{jenkins2007} Jenkins, L.P., Hornschemeier, A.E., Mobasher, M., Alexander, D.M., \& Bauer, F. 2007, \apj, 666, 846

\bibitem[Kavelaars et al.(2000)]{kavelaars2000} Kavelaars, J.J., Harris, W.E., Hanes, D.A., Hesser, J.E., \& Pritchet, C.J. 2000, \apj, 533, 125

\bibitem[Kim et al.(1994)]{kim1994} Kim, K.-T., Kronberg, P.P., Dewdney, P.E., \& Landecker, T.L. 1994, \aaps, 105, 385

\bibitem[Komiyama et al.(2002)]{komiyama2002} Komiyama, Y., Sekiguchi, M., Kashikawa, N., Yagi, M., Doi, M., Iye, M., Okamura, S., Shimasaku, K., Yasuda, N., Mobasher, B., Carter, D., Bridges, T.J., \& Poggianti, B.M. 2002, \apjs, 138, 265

\bibitem[Ledlow \& Owen(1996)]{ledlow1996} Ledlow, M.J., \& Owen, F.N. 1996, \aj, 112, 9

\bibitem[L\'opez-Cruz et al.(2004)]{lopezcruz2004} L\'opez-Cruz, O., Barkhouse, W.A., \& Yee, H.K.C. 2004, \apj, 614, 679

\bibitem[Miller \& Owen(2002)]{miller2002} Miller, N.A., \& Owen, F.N. 2002, \aj, 124, 2453

\bibitem[Miller et al.(2008)]{rlf} Miller, N.A., Hornschemeier, A.E., \& Mobasher, B., Bridges, T.J., Hudson, M.J., Marzke, R.O., \& Smith, R.J. 2008, \aj, in submission

\bibitem[Mobasher et al.(2001)]{mobasher2001} Mobasher, B. et al. 2001, \apjs, 137, 279

\bibitem[Mobasher et al.(2003)]{mobasher2003} Mobasher, B. et al. 2003, \apj, 587, 605

\bibitem[Neumann et al.(2003)]{neumann2003} Neumann, D.M., Lumb, D.H., Pratt, G.W., \& Briel, U.G. 2003, \aap, 400, 811

\bibitem[Oke \& Gunn(1983)]{oke1983} Oke, J.B., \& Gunn, J.E. 1983, \apj, 266, 713

\bibitem[Poggianti et al.(2001a)]{poggianti2001a} Poggianti, B.M., Bridges, T.J., Mobasher, B., Carter, D., Doi, M., Iye, M., Kashikawa, N., Komiyama, Y., Okamura, S., Sekiguchi, M., Shimasaku, K., Yagi, M., \& Yasuda, N. 2001a, \apj, 562, 689

\bibitem[Poggianti et al.(2001b)]{poggianti2001b} Poggianti, B.M., Bridges, T.J., Carter, D., Mobasher, B., Doi, M., Iye, M., Kashikawa, N., Komiyama, Y., Okamura, S., Sekiguchi, M., Shimasaku, K., Yagi, M., \& Yasuda, N. 2001b, \apj, 563, 118

\bibitem[Poggianti et al.(2004)]{poggianti2004} Poggianti, B.M., Bridges, T.J., Komiyama, Y., Yagi, M., Carter, D., Mobasher, B., Okamura, S., \& Kashikawa, N. 2004, \apj, 601, 197

\bibitem[Simpson et al.(2006)]{simpson2006} Simpson, C., Mart\'inez-Sansigre, A., Rawlings, S., Ivison, R., Akiyama, M., Sekiguchi, K., Takata, T., Ueda, Y., \& Watson, M. 2006, \mnras, 372, 741

\bibitem[Spergel et al.(2007)]{spergel2007} Spergel, D.N. et al. 2007, \apjs, 170, 377

\bibitem[Stoughton et al.(2002)]{stoughton2002} Stoughton, C. et al. 2002, \aj, 123, 485 (SDSS EDR)

\bibitem[Struble \& Rood(1999)]{struble1999} Struble, M.F., \& Rood, H.J. 1999, \apjs, 125, 35

\bibitem[Tully \& Pierce(2000)]{tully2000} Tully, R.B., \& Pierce, M.J. 2000, \apj, 533, 744

\bibitem[Vikhlinin et al.(2001)]{vikhlinin2001} Vikhlinin, A., Markevitch, M., Forman, W., \& Jones, C. 2001, \apj, 555, L87 

\bibitem[White et al.(1993)]{white1993} White, S.D.M., Briel, U.G., \& Henry, J.P. 1993, \mnras, 261, L8

\bibitem[York et al.(2000)]{york2000} York, D.G. et al. 2000, \aj, 120, 1579 (SDSS)

\bibitem[Yun, Reddy, \& Condon(2001)]{yun2001} Yun, M.S., Reddy, N.A., \& Condon, J.J. 2001, \apj, 554, 803

\end{thebibliography}
\end{document}